\documentclass[times,twocolumn,final]{elsarticle}
\usepackage{medima}
\usepackage{framed,multirow}
\usepackage{amssymb}
\usepackage{latexsym}
\usepackage{url}
\usepackage{xcolor}

\usepackage{hyperref}

\usepackage{amsmath,amssymb,amsfonts}
\usepackage{algorithmic}
\usepackage{graphicx}
\usepackage{textcomp}
\usepackage{mathrsfs}
\usepackage{placeins}
\usepackage{subfigure}
\usepackage{booktabs}
\usepackage{multirow}
\usepackage{float}
\usepackage{wrapfig}
\usepackage{bbding}
\usepackage{caption}
\usepackage{subcaption}
\usepackage{algorithm}
\usepackage{algorithmic}
\usepackage{lineno}

\definecolor{newcolor}{rgb}{.8,.349,.1}

\journal{Medical Image Analysis}

\begin{document}

\verso{Given-name Surname \textit{et~al.}}

\begin{frontmatter}

\title{RSFR: A Coarse-to-Fine Reconstruction Framework for Diffusion Tensor Cardiac MRI with Semantic-Aware Refinement}%
% \tnotetext[tnote1]{This is an example for title footnote coding.}

\author[1,2,3]{Jiahao \snm{Huang}}
\author[1,2,3]{Fanwen \snm{Wang}}
\author[2,3]{Pedro F. \snm{Ferreira}}
\author[1]{Haosen \snm{Zhang}}
\author[1,2,3]{Yinzhe \snm{Wu}}
\author[4]{Zhifan \snm{Gao}}
\author[5]{Lei \snm{Zhu}}
\author[6]{Angelica I. \snm{Aviles-Rivero}}
\author[7]{Carola-Bibiane \snm{Sch{\"o}nlieb}}
\author[2,3]{Andrew D. \snm{Scott}}
\author[2,3]{Zohya \snm{Khalique}}
\author[2,3]{Maria \snm{Dwornik}}
\author[2,3]{Ramyah \snm{Rajakulasingam}}
\author[2,3]{Ranil De \snm{Silva}}
\author[2,3]{Dudley J. \snm{Pennell}}
\author[1,2,3,8]{Guang \snm{Yang}\corref{cor1}\fnref{fn1}}
\ead{g.yang@imperial.ac.uk}
\author[2,3]{Sonia \snm{Nielles‐Vallespin}\fnref{fn1}}
\cortext[cor1]{Corresponding author: Guang Yang}
\fntext[fn1]{Co-senior author: Guang Yang, Sonia Nielles‐Vallespin}

\address[1]{Department of Bioengineering and Imperial-X, Imperial College London, London, United Kingdom}
\address[2]{National Heart and Lung Institute, Imperial College London, London, United Kingdom}
\address[3]{Cardiovascular Research Centre, Royal Brompton Hospital, London, United Kingdom}
\address[4]{School of Biomedical Engineering, Shenzhen Campus of Sun Yat-sen University, Guangzhou, China}
\address[5]{Robotics and Autonomous Systems Thrust \& Data Science and Analytics Thrust, HKUST (GZ), Guangzhou, China}
\address[6]{Yau Mathematical Sciences Centre, Tsinghua University, Beijing, China.}
\address[7]{Department of Applied Mathematics and Theoretical Physics, University of Cambridge, Cambridge, United Kingdom}
\address[8]{School of Biomedical Engineering and Imaging Sciences, King’s College London, London, United Kingdom}

% \received{1 May 2013}
% \finalform{10 May 2013}
% \accepted{13 May 2013}
% \availableonline{15 May 2013}
% \communicated{S. Sarkar}

\begin{abstract}
Cardiac diffusion tensor imaging (DTI) offers unique insights into cardiomyocyte arrangements, bridging the gap between microscopic and macroscopic cardiac function. 
However, its clinical utility is limited by technical challenges, including a low signal-to-noise ratio, aliasing artefacts, and the need for accurate quantitative fidelity.
To address these limitations, we introduce RSFR (Reconstruction, Segmentation, Fusion \& Refinement), a novel framework for cardiac diffusion-weighted image reconstruction. 
RSFR employs a coarse-to-fine strategy, leveraging zero-shot semantic priors via the Segment Anything Model and a robust Vision Mamba-based reconstruction backbone.
Our framework integrates semantic features effectively to mitigate artefacts and enhance fidelity, achieving state-of-the-art reconstruction quality and accurate DT parameter estimation under high undersampling rates.
Extensive experiments and ablation studies demonstrate the superior performance of RSFR compared to existing methods, highlighting its robustness, scalability, and potential for clinical translation in quantitative cardiac DTI.
\end{abstract}


\end{frontmatter}

%\linenumbers

\section{Introduction}
\label{sec:introduction}

% Advantage of cDTI
Cardiac diffusion tensor imaging (cDTI) is an advanced Magnetic Resonance Imaging (MRI) technique capable of characterising the in vivo myocardial microstructure. 
Water diffusion in myocardial tissue is anisotropic due to its microstructure, and this can be modelled using three-dimensional (3D) tensors with ellipsoid shapes and orientations. 
cDTI offers valuable insights into the myocardial microstructure, bridging the gap between cellular contraction and macroscopic cardiac function~\cite{Ferreira2014Invivo,Sonia2017Assessment}. It also holds potential for assessing myocardial health and developing novel therapeutic strategies~\cite{Khalique2020Diffusion}. Early clinical applications, such as studies on cardiomyopathy and myocardial infarction, have shown promising results, underscoring the potential of cDTI in clinical practice~\cite{Sharrack2022Relationship}.

% Disadvantage of cDTI
While cDTI offers significant advantages, technical challenges hinder its application to routine clinical practice. Calculating the DT requires diffusion-weighted images (DWIs) encoded in at least six non-collinear directions. 
To address the bulk motion from cardiac and respiratory activity, in vivo cDTI employs rapid single-shot acquisitions, such as single-shot echo planar imaging (SS-EPI) or spiral diffusion-weighted imaging. However, these methods produce low signal-to-noise ratio (SNR) images, necessitating multiple repetitions to improve DT estimation accuracy~\cite{Scott2016Effects}. 
Breath-hold acquisitions further complicate this process, as each repetition requires an additional breath-hold, increasing scan time, and causing patient discomfort.
% Accelerate cDTI
Deep learning-based MRI reconstruction has achieved significant advancements by leveraging large MRI datasets to learn complex and hierarchical representations~\cite{Huang2024Data}, which have been employed to accelerate cDTI~\cite{Phipps2021Accelerated,Huang2024Deep}.

% Deep learning has been extensively applied to MRI reconstruction, with four representative approaches~\cite{Hammernik2022Physics}:  
% 1) algorithm unrolling models~\cite{Schlemper2017DCCNN,Aggarwal2019MoDL}, 
% 2) plug-and-play-based models~\cite{Yazdanpanah2019Deep, Ahmad2020Plug}, 
% 3) image enhancement-based models~\cite{Hyun2018Deep, Huang2022SwinMR}, 
% and 
% 4) emerging generative models~\cite{Yang2018DAGAN, Chung2022Score}. 
% The proposed RSFR can be categorised into the image enhancement-based models.

\begin{figure*}[htbp]
    \centering
    \includegraphics[width=\linewidth]{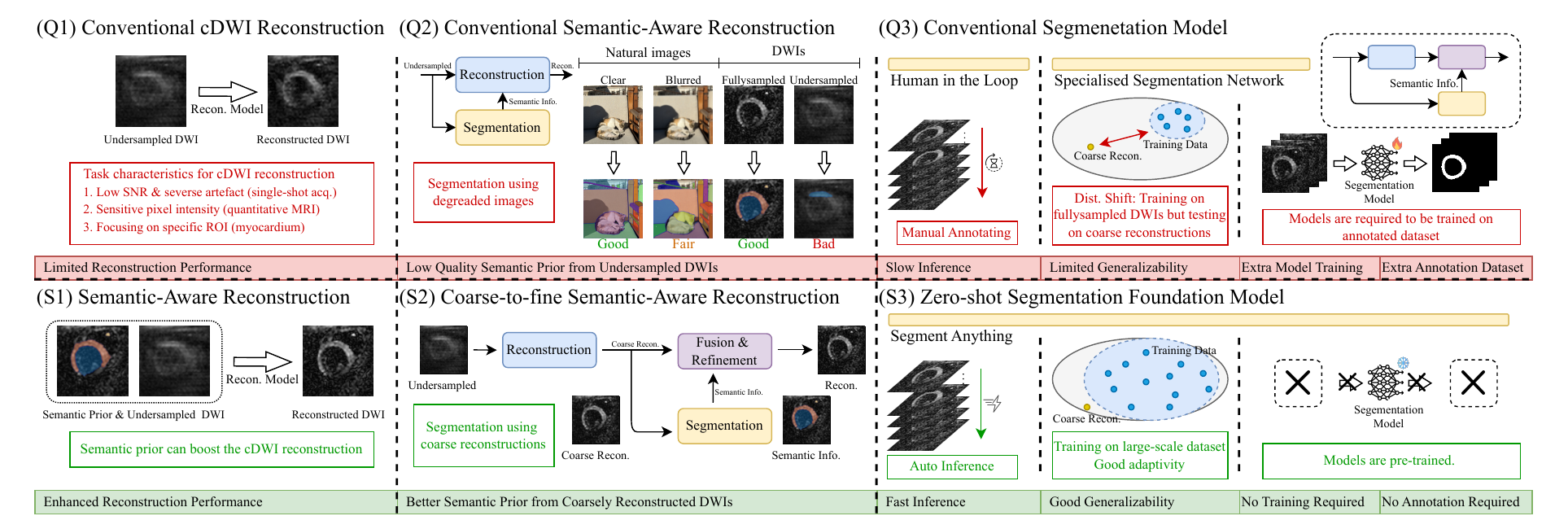}
    \caption{
    The contribution of our RSFR framework: 
    (Q1) {Conventional cDWI reconstruction} suffers from low SNR, severe aliasing artefacts, and ROI-specific intensity sensitivity, resulting in limited reconstruction performance.
    (S1) \textbf{Semantic-aware reconstruction} integrates semantic priors into the reconstruction process, enhancing image quality and fidelity. 
    (Q2) {Conventional semantic-aware reconstruction} extracts semantic priors directly from undersampled DWIs, which is unreliable due to poor segmentation quality from degraded images. 
    (S2) \textbf{Coarse-to-fine semantic-aware reconstruction} improves segmentation reliability by generating semantic priors from coarse reconstructions, leading to better reconstruction performance. 
    (Q3) {Conventional segmentation models} rely on labour-intensive manual annotation or specialised models, which suffer from distribution shifts and require additional training datasets.
    (S3) \textbf{Zero-shot segmentation foundation model} eliminates the need for manual annotation and task-specific model training, improving generalizability. 
    }
    \label{fig:FIG_INTRO}
\end{figure*}

% Q1: Can we mitigate the aliasing artefact from non-ROI?
% S1: Semantic Prior-aware Recon!
Despite these advancements, DWI reconstruction, particularly for cDTI applications, presents unique challenges (Fig.~\ref{fig:FIG_INTRO} {Q1}). A critical issue arises from the inherent low SNR of DWIs, primarily caused by signal attenuation during the encoding of water molecule diffusion~\cite{Ferreira2022Accelerating}.
Furthermore, cardiac DWI demands high pixel intensity accuracy in the myocardium, as precise DT estimation relies on subtle quantitative variations~\cite{Ferreira2014Invivo, Sonia2017Assessment}.
This sensitivity highlights the need for robust reconstruction methods capable of preserving subtle microstructural details and local contrasts while effectively mitigating noise and artefacts.
% A potential solution is to adopt a semantic-aware reconstruction strategy (Fig.~\ref{fig:FIG_INTRO} {S1}), ensuring that the reconstruction process prioritises the accuracy and quality of data within the ROI while suppressing non-ROI contributions.
One promising approach is to adopt a semantic-aware reconstruction strategy (Fig.~\ref{fig:FIG_INTRO} {S1}), as it enables the model to prioritise the accuracy and quality of data within the ROI while suppressing non-ROI contributions. Unlike conventional intensity-based reconstruction, which treats all pixels equally, a semantic-aware approach can leverage prior knowledge about myocardial structure to guide the reconstruction process. This strategy is particularly appealing but challenging, since it aligns with the clinical objective of improving myocardial DTI accuracy but its effectiveness depends on the availability and reliability of extracted semantic priors. 

% Q2: Difficult direct acquire semantic prior from undersampled DWI.
% S2: Coarse-to-fine Reconstruction. semantic prior can be acquired from the coarse reconstruction.
However, acquiring high-quality semantic priors from undersampled DWI presents a critical challenge due to the severe loss of structural and contextual information in highly undersampled images. This degradation makes it difficult to directly extract meaningful semantic priors, as crucial anatomical details required for accurate segmentation may be lost.
Most existing semantic-aware image restoration methods derive semantic priors directly from degraded images~\cite{Xiao2023Dive, Jiang2023Restore} (Fig.\ref{fig:FIG_INTRO} {Q2}), which may be adequate for natural images but are ineffective for undersampled DWIs. 
Zero-filled reconstructions from undersampled DWIs often lack sufficient structural integrity, rendering direct segmentation unreliable, whether using pre-trained segmentation networks or manual annotation by clinicians. 
A promising strategy involves adopting a coarse-to-fine reconstruction framework (Fig.\ref{fig:FIG_INTRO} S2), where an initial coarse reconstruction restores enough anatomical context to enable segmentation and extraction of semantic features, which in turn guide a subsequent, more refined reconstruction process.

% We develop RSFR framework: Reconstruct, Segment, Fuse \& Refine.
To address these challenges, we propose a novel cardiac DWI reconstruction framework, namely RSFR (Reconstruct, Segment, Fuse \& Refine) for cDTI, which integrates advanced deep learning techniques across three key modules:

% Reconstruction Model
\textbf{Reconstruction Model.}
One of the foundational components of the RSFR framework is the reconstruction model.
To mitigate the impact of poor image integrity in undersampled DWIs, we develop an advanced Mamba-based reconstruction backbone to enhance initial reconstructions by recovering fine structural details and reducing noise, making its output more suitable for the subsequent segmentation.
By ensuring a structurally reliable coarse reconstruction, this model addresses the challenge from highly degraded undersampled DWIs, provides a high-quality input for segmentation, ultimately improving the accuracy and robustness of the entire reconstruction pipeline.

% Fusion \& Refinement Model
\textbf{Fusion \& Refinement Model.}
After coarse reconstruction and initial segmentation, the fusion and refinement model addresses the challenge of effectively integrating semantic priors into the reconstruction process.
We introduce a novel Semantic Feature Integration (SFI) module, which combines Segment Anything Model (SAM) derived semantic priors with the reconstructed image data to enhance fine structural details and suppress non-ROI contributions.
By integrating anatomical knowledge into the refinement process, this model enhances myocardial region consistency and improves the overall reconstruction accuracy.

% Segmentation Model: Q3 \& S3
\textbf{Segmentation Model.}
The segmentation model plays a critical role in the RSFR pipeline, as accurate and efficient semantic feature segmentation directly impacts the quality of the refined reconstruction (Fig.~\ref{fig:FIG_INTRO} {Q3}). However, obtaining reliable myocardial segmentation in undersampled DWIs is particularly challenging due to severe aliasing artefacts, low SNR, and misalignment, which collectively degrade structural integrity and hinder accurate feature extraction. 
Conventional segmentation methods typically fall into two categories: manual intervention and specialised deep learning models, both of which have significant limitations. 
% Human in the Loop
Manual segmentation, while accurate, is inherently slow and labour-intensive, as a single DT slice requires more than 70 misaligned DWIs to be manually annotated. This process is not only inefficient but also introduces inter-observer variability and subjectivity.
% Specialized segmentation models
On the other hand, specialised segmentation models, though computationally efficient, often struggle with distribution shifts. These models are typically trained on high-quality ground truth DWIs but perform poorly when applied to noisy, undersampled reconstructions, leading to reduced generalisation. Moreover, they require large training datasets, model tuning, and retraining whenever variations in acquisition conditions arise, significantly increasing the development burden.
% Zero-shot Segmentation Models
To overcome these challenges, we adopt a zero-shot segmentation approach using the Segment Anything Model (SAM)\cite{Kirillov2023Segment} for semantic prior acquisition (Fig.\ref{fig:FIG_INTRO} {S3}). 
Unlike conventional specialised segmentation models that rely on large annotated datasets and task-specific training, SAM can extract rich semantic information directly from undersampled images without requiring prior knowledge of specific data distributions. Instead of generating precise myocardial segmentation masks, SAM provides a high-level semantic prior, which is later refined during the fusion and refinement stage to enhance reconstruction quality.
% Overall contribution.
By leveraging SAM's zero-shot capability, our segmentation model effectively addresses the challenge of unreliable segmentation in undersampled DWI data. 
This approach enables robust myocardial feature extraction without dataset-specific training, enhances generalisation across varying imaging conditions, and seamlessly integrates into the RSFR framework to improve the fidelity of the final reconstruction. These improvements make RSFR more scalable, efficient, and clinically applicable.

% \begin{figure}[htbp]
%     \centering
%     \includegraphics[width=\linewidth]{FIGURE/FIG_SEG_MASK_V2.pdf}
%     \caption{
%     (A) Ground truth DWI; 
%     (B) Manual annotation for myocardium; 
%     (C) Segmentation from a pre-trained UNet~\cite{Ronneberger2015UNet}; 
%     (D) Zero-shot segmentation with top-3 scores from SAM~\cite{Kirillov2023Segment}.
%     }
%     \label{fig:FIG_SEG_MASK}
% \end{figure}

% Experiment & Ablation Studies
To evaluate the performance of our proposed RSFR framework, comprehensive experiments with different undersampling rates across other state-of-the-art (SOTA) reconstruction methods were conducted. 
Experimental results show that our proposed RSFR achieved SOTA on the assessments of reconstruction quality and DT parameter quality. RSFR is able to recover fine structural details with superior pixel-level accuracy and quantitative consistency across different undersampling rates.
Ablation studies have demonstrated the effectiveness of the design within the RSFR framework, by isolating and assessing the contributions of individual components.

In summary, the proposed RSFR framework offers a comprehensive solution for cardiac DWI reconstruction by introducing a semantic-aware reconstruction strategy, leveraging semantic priors through a novel fusion module, and employing the zero-shot segmentation model for efficient and generalisable segmentation. It significantly improves reconstruction quality, enhances the quantitative accuracy of DT parameters, and integrates an end-to-end pipeline for robust and efficient processing, addressing key limitations of existing methods. 
These contributions advance the feasibility and potential clinical translation of cDTI.

\section{Related Work}
\label{sec:related_work}

\subsection{Semantic-Aware Image Restoration}

% Semantic-Aware Image Restoration
Semantic-aware image restoration is an advanced approach that integrates high-level semantic prior into the image restoration process to enhance the quality and accuracy of the results. Traditional image restoration methods typically enhance degraded images globally and uniformly, often neglecting the semantic information of different regions~\cite{Wu2023Learning}. 
Incorporating semantic information enables more effective handling of complex degradations by allowing the restoration process to adapt to the non-uniform distribution of degradation and noise across different regions of an image. 
% Conventional Limitations
Conventional semantic prior-guided restoration methods~\cite{Wang2018Recovering,Wu2023Learning} rely on pre-trained segmentation models or annotated data, limiting adaptability and scalability. These methods often introduce additional inputs or modifications to the restoration pipeline, increasing complexity and computational overhead, which reduces their efficiency in real-world applications.

% SAM Advantages
Segment Anything~\cite{Kirillov2023Segment}, a zero-shot segmentation foundation model, addresses these limitations with zero-shot segmentation capabilities, eliminating the need for task-specific annotations or fine-tuning. 
% Applications in Image Restoration
Recent works have explored SAM's potential in semantic-aware image restoration. 
Jin~\textit{et~al.}~\cite{Jin2023Let} introduced a SAM-guided image dehazing model, while Xiao~\textit{et~al.}~\cite{Xiao2023Dive} utilised SAM priors to enhance existing restoration networks with parameter-efficient tuning. Jiang~\textit{et~al.}~\cite{Jiang2023Restore} proposed the Restore Anything Pipeline, which integrates a controllable model for per-object-level restoration, offering diverse outcomes for user selection.

% Challenges and Advances
Despite these advances, SAM-based methods often face limitations in extracting optimal semantic priors, particularly when dealing with low-quality inputs. For example, severely degraded inputs, such as undersampled cardiac DWI in our task, result in poor semantic prior extraction and suboptimal performance. To address this issue, Zhang~\textit{et~al.}~\cite{Zhang2024Distilling} proposed a semantic prior distillation framework where SAM priors are derived from coarsely recovered images instead of degraded inputs, improving their effectiveness in image restoration tasks.

\subsection{Mamba for Image Restoration}

% Mamba for Vision
Mamba~\cite{Gu2023Mamba} has emerged as a strong alternative to Transformers, offering global sensitivity with linear computational complexity. However, due to its inherently sequential processing, additional adaptations are required to make Mamba suitable for vision tasks.
Vim~\cite{Zhu2024VisionMamba} introduced a plain Mamba-based vision backbone that adapts Mamba for non-causal visual sequences using bi-directional scans and position embeddings. 
VMamba~\cite{Liu2024VMamba} proposed a hierarchical backbone that employs a cross-scan mechanism to construct four ordered patch sequences by integrating pixels from multiple directions.

% Mamba for IR
In the realm of image restoration, Mamba-based models have been successfully applied to various tasks, including image super-resolution, denoising, and reconstruction.
MambaIR~\cite{Guo2024MambaIR} enhanced the vanilla Vision Mamba by integrating convolution and channel attention mechanisms. Further advancements included MambaIRv2~\cite{Guo2024MambaIRv2}, which introduces non-causal modeling capabilities, enabling an attentive state-space restoration model that efficiently processes long-range dependencies in image data.
CU-Mamba~\cite{Deng2024CUMamba} incorporated a dual State Space Model framework within a U-Net architecture, leveraging both spatial and channel SSM modules. This design captures global context while preserving channel correlation features, leading to superior performance in image restoration tasks.
Additionally, MambaMIR~\cite{Huang2024Enhancing_MambaMIR} introduced a wavelet-based U-shaped Mamba architecture that combines Monte Carlo dropout with a cross-scan mechanism, enabling joint medical image reconstruction and uncertainty estimation.
These developments highlight the versatility and effectiveness of Mamba-based architectures in tackling various image restoration challenges, positioning them as strong alternatives to traditional convolutional and transformer-based models.

\section{Methodology}
\label{sec:methodology}

\subsection{Coarse-to-Fine DWI Reconstruction Framework}

% % Formulation
% The data acquisition process of MRI can be formulated as:
% \begin{equation}\label{eq:mri_forward}
% \begin{aligned}
% \mathbf{y}=\mathbf{A} \mathbf{x}+\mathbf{n},
% \end{aligned}
% \end{equation} 
% \noindent where the degradation matrix $\mathbf{A} \in \mathbb{C}^{m \times n}$ is defined as the combination of the undersampling trajectory $\mathbf{M} \in \mathbb{C}^{m \times n}$, the discrete Fourier transform matrix $\mathcal{F} \in \mathbb{C}^{n \times n}$, and a diagonal matrix representing coil sensitivity maps $\mathbf{S} \in \mathbb{C}^{n \times n}$. 

The goal of MRI reconstruction is to recover the target image $\mathbf{x} \in \mathbb{C}^n$ from undersampled \textit{k}-space measurements $\mathbf{y} \in \mathbb{C}^m$, formulated as an inverse problem:
\begin{equation}\label{eq:mri_reverse}
\begin{aligned}
\mathbf{\hat x} = \operatorname{arg}\min_{\mathbf{x}} \frac{1}{2}|| \mathbf{A} \mathbf{x} - \mathbf{y} ||_2^2 + \lambda \mathcal{R}(\mathbf{x}),
\end{aligned}
\end{equation}
\noindent where $\mathbf{A} \in \mathbb{C}^{m \times n}$ is the degradation matrix and the regularisation term $\mathcal{R}(\mathbf{x})$ is weighted by the coefficient $\lambda$. 

The proposed RSFR framework adopts a coarse-to-fine approach to address the unique challenges of DWI reconstruction in cDTI, by leveraging semantic information, ensuring robust and accurate reconstruction while mitigating artefacts from noisy backgrounds. 
The RSFR begins with the Reconstruction Model $\operatorname{H}_{\operatorname{R}}$, which performs a coarse reconstruction $\mathbf{\bar x}$ on the zero-filled undersampled DWI $\mathbf{A}^{H} \mathbf{y}$. 
This initial reconstruction provides sufficient structural information and serves as the input to the Segmentation Model $\operatorname{H}_{\operatorname{S}}$, which employs zero-shot semantic segmentation to extract the semantic prior $\mathbf{F}_{\text{seg}}$. 
The Fusion \& Refinement Model $\operatorname{H}_{\operatorname{FR}}$ integrates the coarse reconstruction and the semantic information derived from the Segmentation Model, to produce the refined reconstruction $\mathbf{\hat x}$. 
Our proposed RSFR framework can be presented as follows:
\begin{equation}\label{eq:framework}
\begin{aligned}
\mathbf{\bar x} = \operatorname{H}_{\operatorname{R}}(\mathbf{A}^{H} \mathbf{y}), \quad
\mathbf{F}_{\text{seg}} = \operatorname{H}_{\operatorname{S}}(\mathbf{\bar x}), \quad
\mathbf{\hat x} = \operatorname{H}_{\operatorname{FR}}(\mathbf{\bar x}, \mathbf{F}_{\text{seg}}).
\end{aligned}
\end{equation}

The RSFR framework is inherently designed as a plug-in solution that can be seamlessly integrated with various reconstruction backbones. In this study, we utilise the recently proposed VMamba~\cite{Liu2024VMamba} as the reconstruction backbone, leveraging its superior ability to model long-range dependencies and linear computational complexity. Experiments with alternative backbones are presented and analysed later.

\subsection{Vision Mamba-based Reconstruction Backbone}

In cDTI, undersampled DWI is typically highly degraded, presenting significant challenges for coarse reconstruction and consequently impacting the quality of SAM-derived semantic priors. Hence, utilising a robust reconstruction backbone as the Reconstruction Model is crucial. 

In this work, we design a novel Vision Mamba-based Reconstruction Model and a Fusion \& Refinement Model, as illustrated in Fig.~\ref{fig:FIG_ARCH} (A) and (B) correspondingly (skip connections are omitted for clarity), inspired by the vision backbone VMamba~\cite{Liu2024VMamba} and MambaMIR~\cite{Huang2024Enhancing_MambaMIR}

The reconstruction backbone proposed in this study adopts a U-shaped architecture, consisting of an image encoder for patch embedding, an image decoder for patch unembedding, as well as $n$ encoder and decoder residual Mamba blocks, interconnected through corresponding skip connections. 
In both the encoding and decoding paths, each residual Mamba block incorporates two Visual State Space (VSS) blocks, complemented by modules for downsampling and upsampling. 
Within the bottleneck, a single self-attention block is utilised for processing deep features.

\subsubsection{Visual State Space Block}

The design of the VSS block draws inspiration from the architectures of the Mamba block~\cite{Gu2023Mamba} and the VSS block~\cite{Liu2024VMamba}. 

Within the VSS block, the input first undergoes a normalisation layer and is subsequently divided into two distinct processing pathways. 
The first pathway processes the input through a series of layers consisting of a linear layer, a depth-wise 3$\times$3 convolutional layer, a SiLU activation layer, a core S6 block, and another layer normalisation layer. The second pathway applies a linear layer followed by a SiLU activation layer. 
Finally, the outputs of these two pathways are combined through element-wise multiplication, followed by a gating linear layer, to generate the final output of the VSS block.

\subsubsection{S6 Block and Cross Scan Mechanism}

% Challenge on Mamba for Vision
One challenge with using Mamba to process vision data is that S6 processes information in a strictly ordered sequential manner, meaning that data integration is confined to what has been processed in sequence. Although this is beneficial for temporal natural language processing applications, it presents obstacles for computer vision tasks because the data does not inherently follow a sequential order.
To address this issue, several methods have been developed that involve adjusting the order of the visual sequence in different ways, such as bidirectional scan, cross scan and zigzag scan~\cite{Liu2024Vision}.

% S6
We adopt the S6 block's architecture featuring a cross-scan approach~\cite{Liu2024VMamba} to modify Mamba for use with our DWI data. The S6 block comprises three main components: the Scan Expanding module, the S6 module, and the Scan Merging module.
% Scan Expanding module
The Scan Expanding module takes image patches and stretches them across rows or columns, starting either from the upper-left or the lower-right corner, converting one image into four unique ordered sequences to mitigate the sequential processing issue.
% S6 module
The S6 module is the core operation within the S6 block, tasked with handling scan-expanded sequences. 
% Scan Merging module
Subsequently, the Scan Merging module combines and reassembles these processed scans back into their original patch format.

\begin{figure}[t!]
    \centering
    \includegraphics[width=\linewidth]{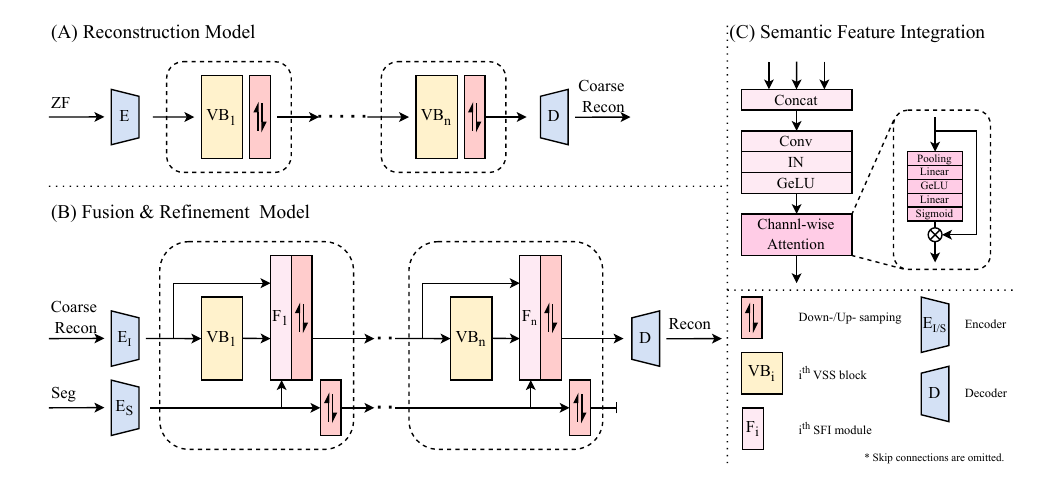}
    \caption{
    The architecture of (A) the Reconstruction Model, (B) the Fusion \& Refinement Model, and (C) the Semantic Feature Integration module.
    }
    \label{fig:FIG_ARCH}
\end{figure}

\subsection{Leverage SAM Semantic Prior for Reconstruction}

% Segmentation Model
The Segmentation Model takes the coarse reconstruction $\hat{\mathbf{x}}{\operatorname{coarse}}$ as input and generates the semantic prior for subsequent refinement. We employ SAM~\cite{Kirillov2023Segment} as the Segmentation Model, leveraging its zero-shot segmentation predictions, where the top three predictions with the highest scores are further utilised as the semantic prior.

% Fusion \& Refinement Model
The novel Fusion \& Refinement Model (Fig.~\ref{fig:FIG_ARCH}(B)) is designed to integrate the coarse reconstruction $\hat{\mathbf{x}}_{\operatorname{coarse}}$ and the semantic priors $\hat{\mathbf{m}}_{\operatorname{sam}}$ derived from the Segmentation Model. This integration aims to generate a refined reconstruction $\hat{\mathbf{x}}$, which not only improves image quality but also preserves critical structural details, effectively addressing the challenges posed by severely degraded undersampled DWI data.

The Fusion \& Refinement Model is constructed using a Vision Mamba-based backbone architecture, similar to that of the Reconstruction Model, but incorporates the specialised Semantic Feature Integration (SFI) module (Fig.~\ref{fig:FIG_ARCH}(C)) to facilitate the efficient fusion of semantic information and suppression of irrelevant details.
The SFI module operates by first concatenating features along the channel dimension, followed by sequential processing through a convolutional layer, an instance normalisation (IN) layer, and a GeLU activation layer to achieve feature alignment. Subsequently, a channel-wise attention mechanism refines the fused features by computing and applying channel-wise attention weights. This selective enhancement ensures that semantic priors effectively guide the refinement process, enabling the Fusion \& Refinement Model to produce high-quality reconstructions while preserving anatomical details, a feature particularly critical for the DWI reconstruction task.

\subsection{Data Flow for cDTI Reconstruction}

\subsubsection{Data Pre-processing}

During the data pre-processing phase, all DWIs (\textit{b0}, \textit{b150}, and \textit{b600}) underwent the same processing protocol.

% Normalisation
All DWIs were normalised to a $0 \sim 1$ range with the max-min method slice by slice, due to the significant pixel value discrepancies across various \textit{b}-values. Meanwhile, all the maximum and minimum pixel values were utilised to restore the original pixel intensity range in the post-processing stage.
% Zero-padding 
Zero-padding was then applied along the phase encoding direction to standardise the resolution to $256 \times 96$.
% Simulation K-space Undersampling
To simulate the \textit{k}-space undersampling process, GRAPPA-like Cartesian \textit{k}-space undersampling masks at AFs$\times 2$, $\times 4$, and $\times 8$ were generated following the official protocol of the fastMRI dataset~\cite{Zbontar2018fastMRI}. 
Since all these retrospectively acquired data were reconstructed with a zero-padding factor of two in the scanner, we employed a phase encoding dimension of 48 instead of 96, to ensure a more realistic simulation. These undersampling masks were subsequently zero-padded from $128 \times 48$ to $256 \times 96$.
% Center Cropping
Before being processed by the RSFR framework, the DWIs were cropped to a resolution of $96 \times 96$ for computational efficiency, as the ROI is located in the central area.

\subsubsection{Data Post-processing}

We utilised our locally developed open-source software, i.e., INDI~\footnote{https://github.com/ImperialCollegeLondon/INDI/tree/main}, for cDTI post-processing, following the protocol described in~\cite{Ferreira2014Invivo,Ferreira2022Accelerating}, which involved the following steps: 1) manual low-quality DWIs discarding; 2) low-rank groupwise DWIs registration~\cite{Wang2024Low}; 3) LV myocardium segmentation; 4) least-squares-based DT estimation; and 5) DT parameter calculation. 
The output of our RSFR framework was restored to its original pixel value range and resolution prior to the post-processing stage.

\subsection{Optimisation Scheme}

The proposed RSFR framework can be trained and tested in an end-to-end manner, where the weights of the Reconstruction Model and the Fusion \& Refinement Model are updated simultaneously, while the weights of SAM remain frozen.

We employed a hybrid loss function, denoted as $\mathcal{L}_{\mathrm{RSFR}}(\theta)$, during training. This loss incorporates the Charbonnier loss applied in both the image space and the \textit{k}-space, represented as $\mathcal{L}_{\mathrm{i}}(\theta)$ and $\mathcal{L}_{\mathrm{k}}(\theta)$, respectively. 
To enhance perceptual reconstruction quality, we imposed an $l_{1}$ constraint on the latent space using a pre-trained VGG model, $f_{\mathrm{VGG}}(\cdot)$, resulting in perceptual loss $\mathcal{L}_{\mathrm{p}}(\theta)$.
These loss functions are defined as:
\begin{equation}\label{eq:loss}
\begin{aligned}
\mathop{\text{min}}\limits_{\theta} \mathcal{L}_{\mathrm{i}}(\theta) 
&=\sqrt{\mid\mid \mathbf{x} - \hat{\mathbf{x}}_u \mid\mid^2_2 + \epsilon^2}, \\
\mathop{\text{min}}\limits_{\theta} \mathcal{L}_{\mathrm{k}}(\theta) 
&=\sqrt{\mid\mid \mathcal{F}\mathbf{x} - \mathcal{F} \hat{\mathbf{x}}_u \mid\mid^2_2 + \epsilon^2}, \\
\mathop{\text{min}}\limits_{\theta} \mathcal{L}_{\mathrm{p}}(\theta) 
&= \mid\mid f_{\mathrm{VGG}}(\mathbf{x}) - f_{\mathrm{VGG}}(\hat{\mathbf{x}}_u) \mid\mid_1, \\
\mathcal{L}_{\mathrm{RSFR}}(\theta)
& = \alpha \mathcal{L}_{\mathrm{i}}(\theta)
+ \beta \mathcal{L}_{\mathrm{k}}(\theta)
+ \gamma \mathcal{L}_{\mathrm{p}}(\theta),
\end{aligned}
\end{equation} 
where $\mathbf{x}$ and $\hat{\mathbf{x}}_u$ denote the ground truth and reconstructed DWIs.
$\theta$ represents the trainable parameters of the proposed RSFR. 
$\epsilon$ in the Charbonnier loss is fixed at $10^{-9}$. 
$\mathcal{F}$ stands for the Discrete Fourier Transform. The parameters $\alpha$, $\beta$, and $\gamma$ are responsible for weighting various loss terms.

\section{Experiments}
\label{sec:experiments}

\subsection{Dataset}

% Dataset Info
% All data utilised in this work were approved by the National Research Ethics Committee, Bloomsbury, under reference number 13/LO/1830, REC reference 10/H0701/112, and IRAS reference number 33773. The study complies with the principles outlined in the Declaration of Helsinki and the UK Research Governance Framework version 2. Informed written consent was obtained from all participants.

% Acuisition Info
Retrospectively acquired cDTI data were obtained using Siemens Skyra 3T and Vida 3T MRI scanners (Siemens AG, Erlangen, Germany). The data acquisition employed a diffusion-weighted stimulated echo acquisition mode (STEAM)-EPI sequence with a reduced phase field-of-view and fat saturation.
DWIs were acquired in six directions with diffusion weightings of $\text{b} = 150 \ \text{and} \ 600 \ \text{s/mm}^\text{2}$ (referred to as \textit{b150} and \textit{b600}) on a short-axis mid-ventricular slice. In addition, reference images (\textit{b0}) were obtained with a minor diffusion weighting.
% GRAPPA and mSENSE were utilised with an acceleration factor (AF) of $\times 2$ for the reconstruction.
All DWIs in this study were retrospectively reconstructed by GRAPPA or mSENSE with an acceleration factor (AF) of $\times 2$ in the scanner, and only magnitude data were accessible. 

% Pathology Info
The dataset comprises a total of $457$ cases, including two cardiac phases: diastole ($n = 223$) and systole ($n = 234$), used for the experiments in this study. 
Specifically, the dataset includes:  
$140$ healthy cases, 
$30$ amyloidosis (AMYLOID) cases, 
$46$ dilated cardiomyopathy (DCM) cases, 
$35$ in-recovery DCM (rDCM) cases, 
$37$ hypertrophic cardiomyopathy (HCM) cases, 
$48$ HCM genotype-positive–phenotype-negative (HCM G+P-) cases,  
and $121$ acute myocardial infarction (MI) cases.
% Dataset Division Info
These $457$ cases were divided into a training set ($434$ cases) and a testing set ($23$ cases). During the training phase, a 5-fold cross-validation strategy was employed for the training set.

\subsection{Implementation Details}

% Hyperparameter
Regarding the hyperparameters of the reconstruction backbone, we employed 8 residual Mamba blocks symmetrically positioned within both the encoder and decoder paths. The initial embedding channel was set to 180, with a scaling factor of $\{1, 2, 2, 2\}$ applied from the shallow layers to the deeper layers. 
For the SAM~\cite{Kirillov2023Segment} used in the Segmentation Model, we utilised the official implementation along with the pre-trained weights of the `ViT-H SAM'.

% Training Strategy
Our RSFR model was trained on two NVIDIA A100 GPUs (80GB) using the Adam optimiser over 100,000 gradient steps with a batch size of 8, and tested on an NVIDIA RTX 4090 GPU (24GB). The learning rate was initialised at 0.0002 and reduced by a factor of 0.5 every 20,000 steps after reaching the 50,000$^\text{th}$ step.

\subsection{Evaluation Metrics}

This study evaluated the performance of the RSFR framework through two key aspects: DWI reconstruction quality and DT parameter accuracy.

% DWI Reconstruction
DWI reconstruction quality was assessed using three metrics: Peak Signal-to-Noise Ratio (PSNR), Structural Similarity Index Measure (SSIM), and Learned Perceptual Image Patch Similarity (LPIPS). 
PSNR and SSIM are pixel-based metrics that directly measure pixel-wise accuracy relative to the reference, while LPIPS, a deep feature-based metric, aligns well with human perceptual judgments. Higher PSNR and SSIM, along with lower LPIPS indicate better reconstruction quality.

% DT Params
% MAE of MD, FA, HA
The accuracy of DT parameters, including mean diffusivity (MD), fractional anisotropy (FA), and helix angle (HA) gradient, was evaluated using the mean absolute error (MAE) between estimated and reference values. Lower MAE signifies better agreement with the reference.

% HA Line Profile Analysis
To further assess DT parameter consistency, the HA Gradient Line Profile (HA LP)~\cite{Wang2024Groupwise} was utilised. 
- % reason
In healthy myocardium, the HA typically reflects the gradual change in orientation of myocardial fibres from left-handed near the endocardium to right-handed near the epicardium. 
In MI cases, this pattern is disrupted, with infarcted regions showing reduced right-handed myocytes~\cite{Sharrack2022Relationship}. 
- % HA LP
Radial line profiles of HA were compared using linear regression metrics ($R^2$ and Root Mean Squared Error (RMSE)) to evaluate alignment with the reference. Smaller discrepancies between the reconstructed and reference results indicate higher consistency and fidelity in DT parameter reconstruction.

\subsection{Comparisons with the SOTA}

The proposed RSFR framework was evaluated against baseline and SOTA methods, including: 
the unrolling-based method D5C5~\cite{Schlemper2017DCCNN}, 
GAN-based methods DAGAN~\cite{Yang2018DAGAN} and STGAN~\cite{Huang2022STGAN}, 
image enhancement-based methods UNet~\cite{Ronneberger2015UNet}, SwinMR~\cite{Huang2022SwinMR}, and MambaMIR~\cite{Huang2024Enhancing_MambaMIR}.
We utilised three undersampling rates at AF$\times 2$, $\times 4$ and $\times 8$.

\subsubsection{Assessment on Reconstruction Quality}

\begin{figure}[t!]
    \centering
    \includegraphics[width=\linewidth]{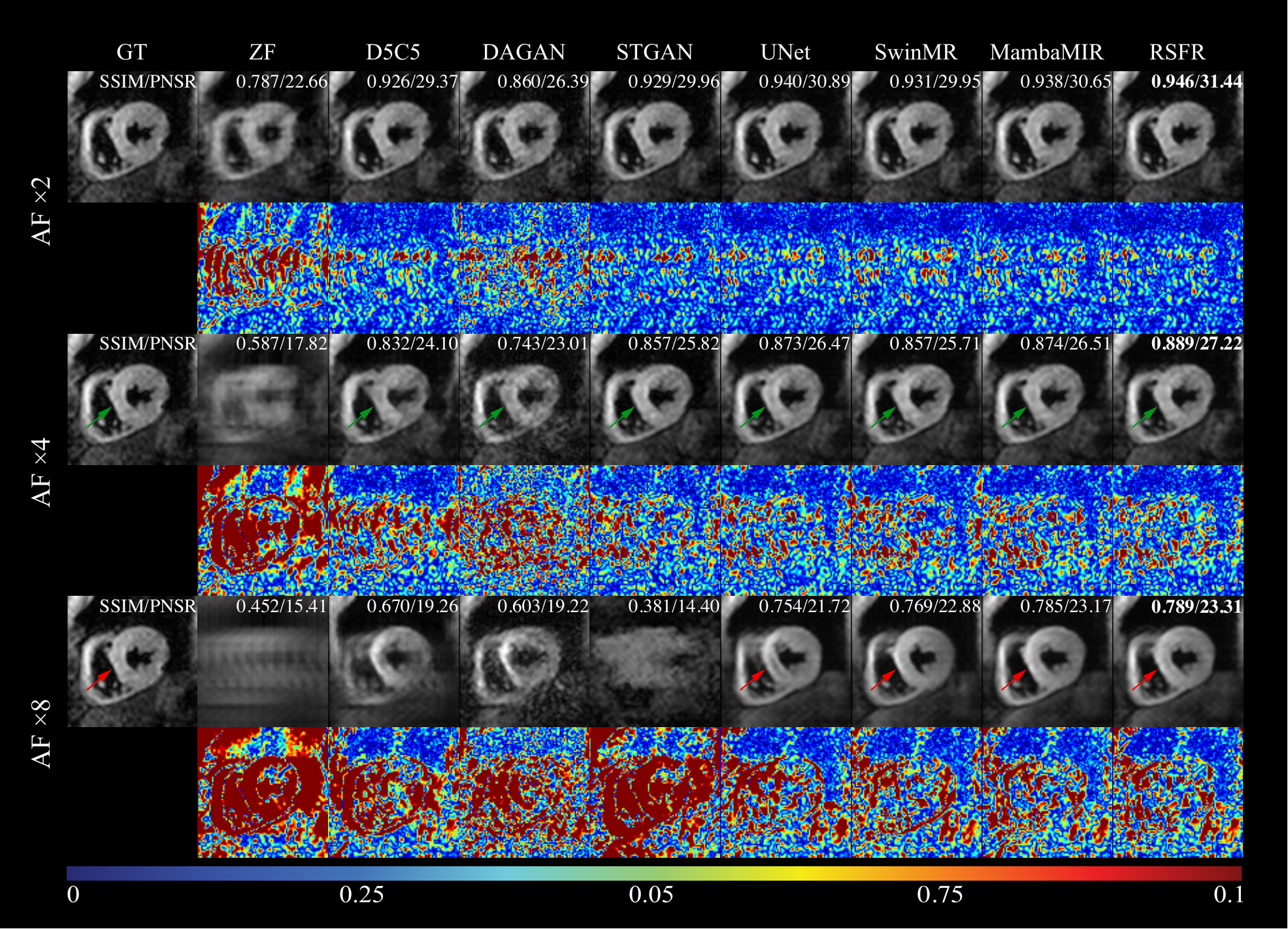}
    \caption{
    Visualised reconstructed DWI samples and corresponding error maps at AF$\times 2$, $\times 4$ and $\times 8$. 
    }
    \label{fig:FIG_RECON_VIS}
\end{figure}

For the assessment on reconstruction quality, the quantitative results and visualised reconstruction samples are presented in TABLE~\ref{tab:TAB_COMP} and Fig.~\ref{fig:FIG_RECON_VIS}, respectively.
The proposed RSFR framework consistently demonstrates superior fidelity performance (SSIM and PSNR) and reasonable perceptual quality (LPIPS) compared to baseline and SOTA methods across all AFs. 
Visualised samples highlight the ability of RSFR to preserve myocardial structure and minimise artefacts, particularly at higher undersampling rates.

\begin{table*}[htbp]
  \centering
  \caption{
  Quantitative results at AF $\times 2$, $\times 4$ and $\times 8$. 
  $^{\star}$ indicates the result is significantly different from the best result by the Mann-Whitney Test ($p<0.05$).
  The \textbf{bold} and \underline{underlined} texts indicate the best and second-best results, respectively.
  }
    \scalebox{0.78}{
    \begin{tabular}{cccccccccc}
    \toprule
          & \multicolumn{3}{c}{AF$\times 2$} & \multicolumn{3}{c}{AF$\times 4$} & \multicolumn{3}{c}{AF$\times 8$} \\
\cmidrule{2-10}    Method & SSIM$\uparrow$ & PSNR$\uparrow$  & LPIPS$\downarrow$ & SSIM$\uparrow$  & PSNR$\uparrow$  & LPIPS$\downarrow$ & SSIM$\uparrow$  & PSNR$\uparrow$  & LPIPS$\downarrow$ \\
    \midrule
    ZF    & 0.825 (0.034)$^{*}$ & 26.58 (2.07)$^{*}$ & 0.146 (0.027)$^{*}$ & 0.676 (0.058)$^{*}$ & 22.13 (2.42)$^{*}$ & 0.326 (0.038)$^{*}$ & 0.551 (0.073)$^{*}$ & 19.56 (2.50)$^{*}$ & 0.490 (0.034)$^{*}$ \\
    D5C5  & 0.924 (0.025)$^{*}$ & 31.64 (2.06)$^{*}$ & 0.058 (0.026)$^{*}$ & 0.849 (0.037)$^{*}$ & 27.69 (1.90)$^{*}$ & 0.099 (0.032)$^{*}$ & 0.701 (0.054)$^{*}$ & 23.03 (2.11)$^{*}$ & 0.202 (0.050)$^{*}$ \\
    DAGAN & 0.867 (0.022)$^{*}$ & 28.74 (1.43)$^{*}$ & \underline{0.047 (0.016)}$^{*}$ & 0.761 (0.038)$^{*}$ & 25.25 (1.76)$^{*}$ & 0.092 (0.028)$^{*}$ & 0.605 (0.055)$^{*}$ & 21.45 (1.91)$^{*}$ & \textbf{0.158 (0.044)} \\
    STGAN & 0.915 (0.028)$^{*}$ & 31.21 (2.14)$^{*}$ & \textbf{0.030 (0.014)} & 0.834 (0.044)$^{*}$ & 27.41 (2.04)$^{*}$ & \textbf{0.058 (0.022)} & 0.457 (0.064)$^{*}$ & 18.17 (1.87)$^{*}$ & 0.381 (0.043)$^{*}$ \\
    UNet  & 0.926 (0.025)$^{*}$ & 31.90 (2.15)$^{*}$ & 0.056 (0.024)$^{*}$ & 0.860 (0.040)$^{*}$ & 28.23 (2.04)$^{*}$ & 0.096 (0.032)$^{*}$ & 0.744 (0.061)$^{*}$ & 24.07 (2.22)$^{*}$ & 0.178 (0.047)$^{*}$ \\
    SwinMR & 0.925 (0.025)$^{*}$ & 31.79 (2.17)$^{*}$ & 0.056 (0.024)$^{*}$ & 0.856 (0.040)$^{*}$ & 28.04 (2.08)$^{*}$ & 0.098 (0.032)$^{*}$ & 0.732 (0.063)$^{*}$ & 23.76 (2.24)$^{*}$ & 0.185 (0.050)$^{*}$ \\
    MambaMIR & 0.929 (0.025)$^{*}$ & 32.09 (2.18)$^{*}$ & 0.055 (0.024)$^{*}$ & 0.864 (0.040)$^{*}$ & 28.45 (2.10)$^{*}$ & 0.095 (0.032)$^{*}$ & 0.748 (0.061)$^{*}$ & 24.18 (2.24)$^{*}$ & 0.175 (0.048)$^{*}$ \\
    \midrule
    RSFR  & \textbf{0.933 (0.023)} & \textbf{32.30 (2.17)} & 0.050 (0.021)$^{*}$ & \textbf{0.871 (0.038)} & \textbf{28.61 (2.07)} & \underline{0.087 (0.028)}$^{*}$ & \textbf{0.754 (0.060)} & \textbf{24.25 (2.21)} & \underline{0.163 (0.041)}$^{*}$ \\
    \bottomrule
    \end{tabular}%
    }
  \label{tab:TAB_COMP}%
  \vspace{-3mm}
\end{table*}%

\subsubsection{Assessment on Diffusion Tensor Parameter}

\begin{figure*}[t!]
    \centering
    \includegraphics[width=\linewidth]{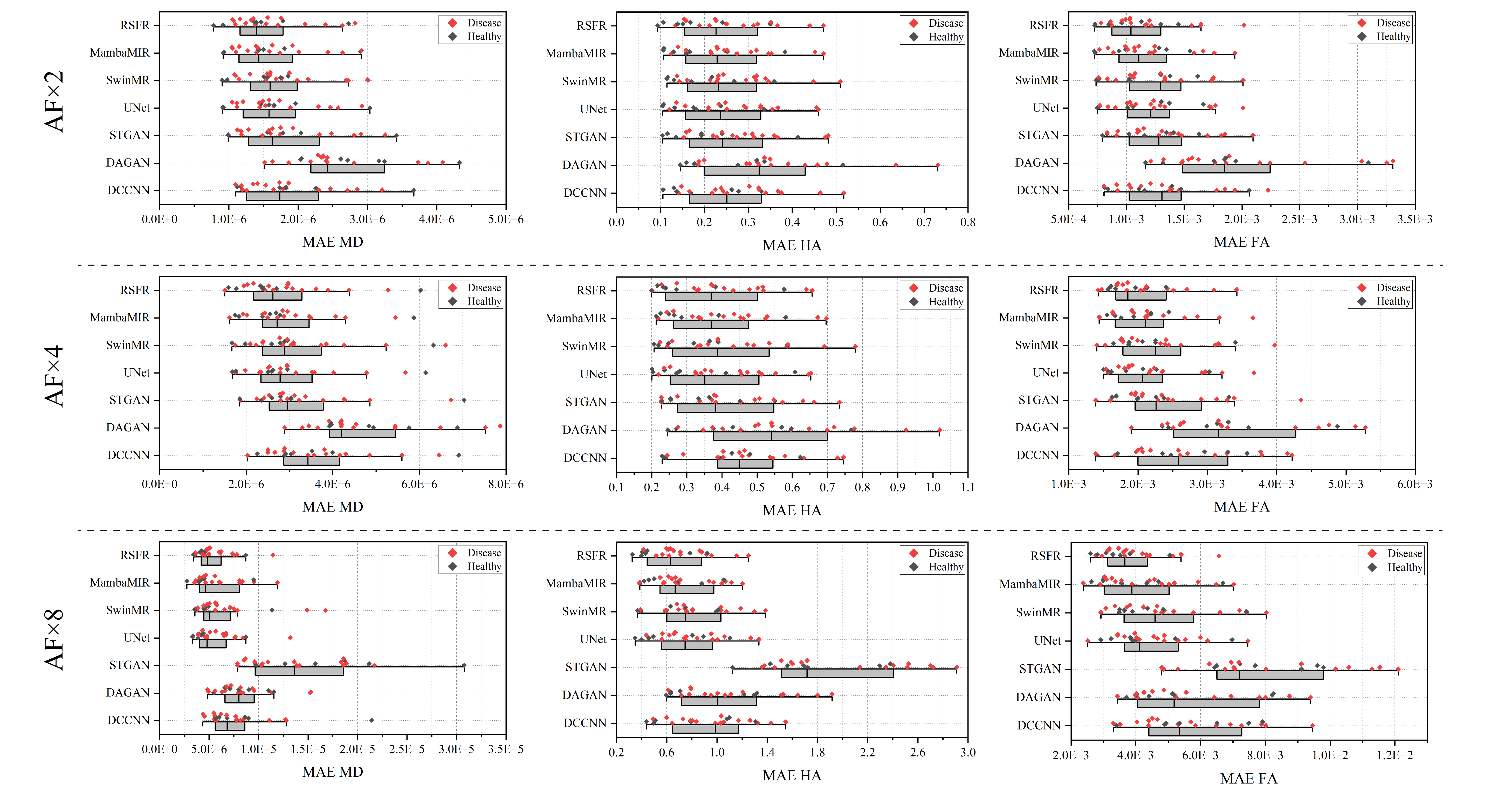}
    \caption{
    Mean absolute error (MAE) of global mean diffusivity (MD), fractional anisotropy (FA), and helix angle (HA) gradient.
    }
    \label{fig:FIG_DT_BP}
\end{figure*}

\begin{figure}[!t]
\begin{subfigure}
    \centering
    \includegraphics[width=\linewidth]{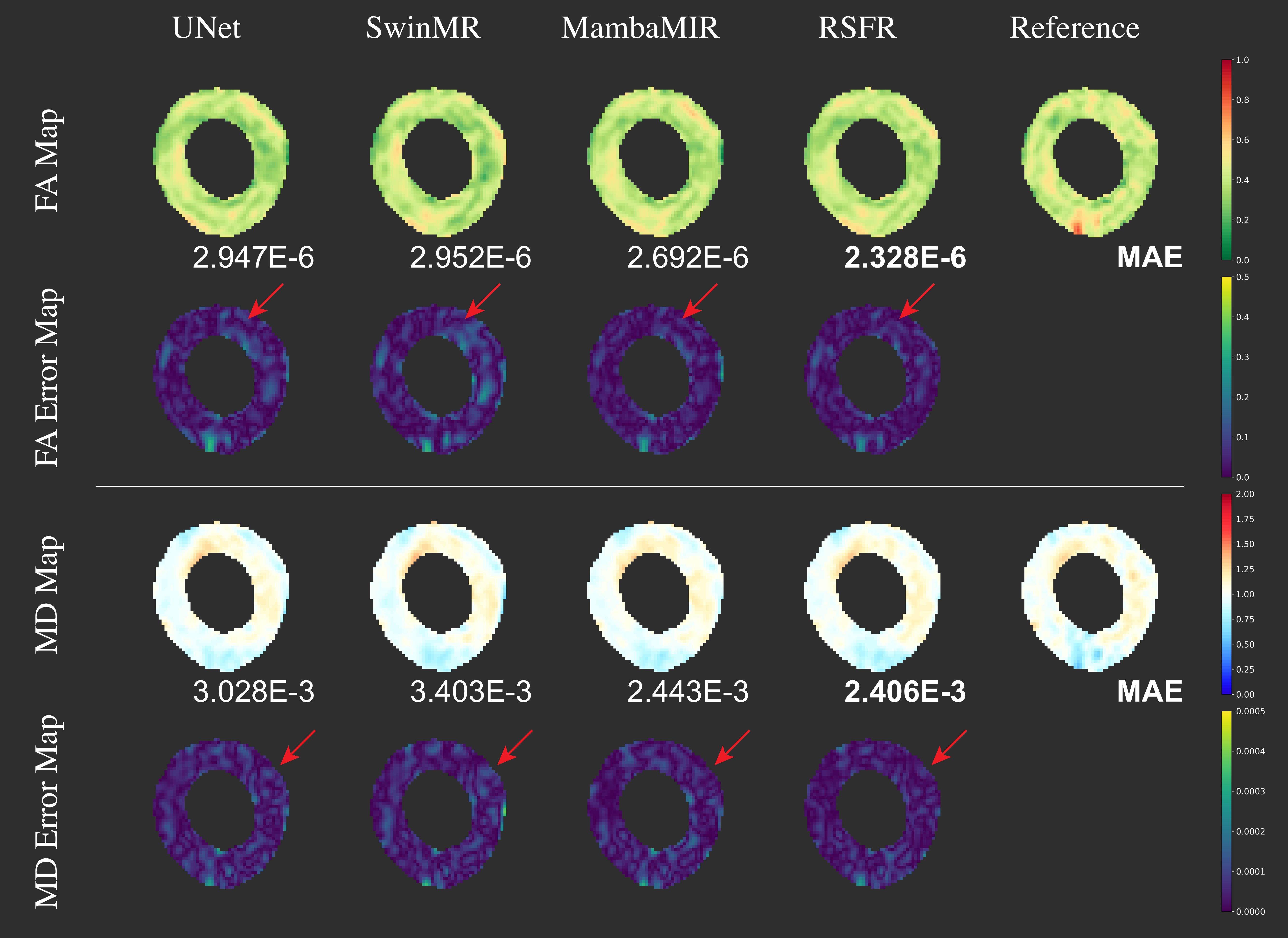}
    \caption{
    Mean diffusivity (MD), fractional anisotropy (FA), and the corresponding error maps for the reference, the proposed RSFR, and selected comparison methods at AF$\times 4$. 
    The red arrows indicate that our RSFR produces the least estimation error. 
    }
    \label{fig:FIG_DT_VIS}
\end{subfigure}

\begin{subfigure}
    \centering
    \includegraphics[width=\linewidth]{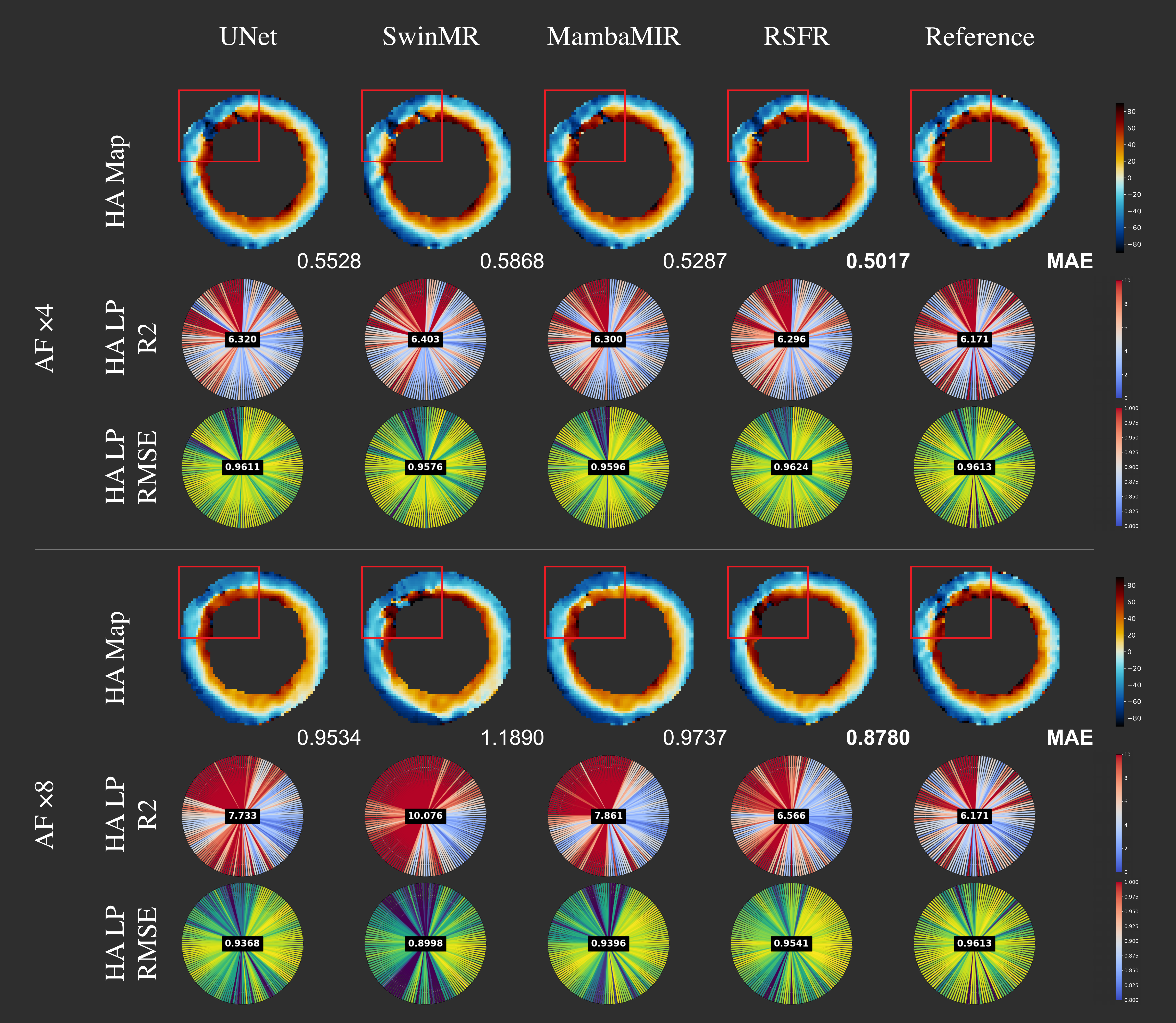}
    \caption{
    Helix angle (HA) maps and HA line profiles (HA LP) for the reference, the proposed RSFR, and selected comparison methods at AF$\times 4$ and AF$\times 8$. 
    The red squares indicate that our RSFR produces the most accuracy estimation in MI area.
    }
    \label{fig:FIG_DT_VIS_MI_ANALYSIS}
\end{subfigure}
\end{figure}

To further evaluate reconstruction performance, post-processing was performed for all comparison methods, and the accuracy of DT parameters, including MD, FA, and HA, was assessed.
Quantitative comparisons in Fig.~\ref{fig:FIG_DT_BP} depict the MAE of global mean MD, FA, and HA gradient for both healthy and diseased cases. RSFR consistently achieved the lowest MAE across nearly all parameters and undersampling rates, demonstrating superior quantitative accuracy. 
Visual comparisons of FA and MD maps, along with their corresponding error maps, are presented in Fig.~\ref{fig:FIG_DT_VIS}. 
RSFR demonstrates the closest alignment to the reference data, with error maps showing minimal deviations, particularly in the myocardial region. 
Fig.~\ref{fig:FIG_DT_VIS_MI_ANALYSIS} presents HA maps and the corresponding HA LP for an MI case. 
RSFR achieves $R^2$ and RMSE values that are closest to reference, accurately replicating the expected transmural variation of HA in the myocardium.

\subsection{Ablation Studies}

Comprehensive ablation studies were conducted to evaluate the contributions of individual components in the RSFR.

\subsubsection{Ablation Studies on Segmentation Model}

% w/ SAM (Ours)
% w/ Ref. Mask: Replace Segmentation Model with the manual annotation.
% w/ UNet: A pre-trained UNet for myocardium segmentation
% N.A.: Remove Segmentation Model

To evaluate the impact of segmentation models on reconstruction performance, we conducted an ablation study comparing SAM-based segmentation (w/ SAM) with configurations using manually annotated reference masks (w/ Ref. Mask), a pre-trained UNet~\cite{Ronneberger2015UNet} segmentation model (w/ UNet), and the absence of segmentation (N.A.). These settings were assessed across AF$\times 2$, $\times 4$ and $\times 8$ using metrics for DWI reconstruction quality (SSIM) and DT parameter quality (MAE of global mean FA), as presented in Fig.~\ref{fig:FIG_ABL_SEG}.

For SSIM, the configuration without segmentation yields the worst performance, while the reference mask configuration (w/ Ref. Mask) consistently achieves the best SSIM scores across all AFs, emphasising the critical role of semantic priors in reconstruction. 
However, the reference mask configuration is impractical in real-world scenarios, as it assumes access to clear annotations before reconstruction. 
The UNet-based configuration performs well at lower AFs but deteriorates significantly at higher AFs, likely due to its susceptibility to distribution shifts since it was trained with ground truth data. 
Additionally, UNet requires a dedicated annotation dataset for training, adding an extra layer of effort. In contrast, the SAM-based approach demonstrates stable and competitive performance across all AFs without requiring extra annotation or training, making it a practical and scalable solution.

For FA errors, the SAM-based configuration consistently outperforms the UNet-based approach, demonstrating superior robustness across all AFs. While the reference mask configuration shows slight advantages at extremely high undersampling rates (AF$\times 8$), SAM achieves comparable or better results in other scenarios, validating its effectiveness in preserving quantitative fidelity without manual annotations or additional datasets.

\subsubsection{Ablation Studies on Reconstruction Backbone}

% Mamba-based + w/ RSFR: The proposed model
% Mamba-based + N.A: Only Reconstruction Model
% SwinMR + w/ RSFR: Use the proposed RSFR framework with SwinMR as the reconstruction backbone,
% SwinMR + N.A: The original SwinMR backbone
% UNet + w/ RSFR: Use the proposed RSFR framework with UNet as the reconstruction backbone,
% UNet + N.A: The original UNet backbone

To evaluate the impact of the reconstruction backbone on the RSFR framework, we tested the proposed Mamba-based backbone, SwinMR~\cite{Huang2022SwinMR}, and UNet~\cite{Ronneberger2015UNet} under two configurations: with RSFR (full framework) and N.A. (only the reconstruction model), across AF$\times 2$, $\times 4$ and $\times 8$).

\begin{figure}[!t]
\begin{subfigure}
    \centering
    \includegraphics[width=\linewidth]{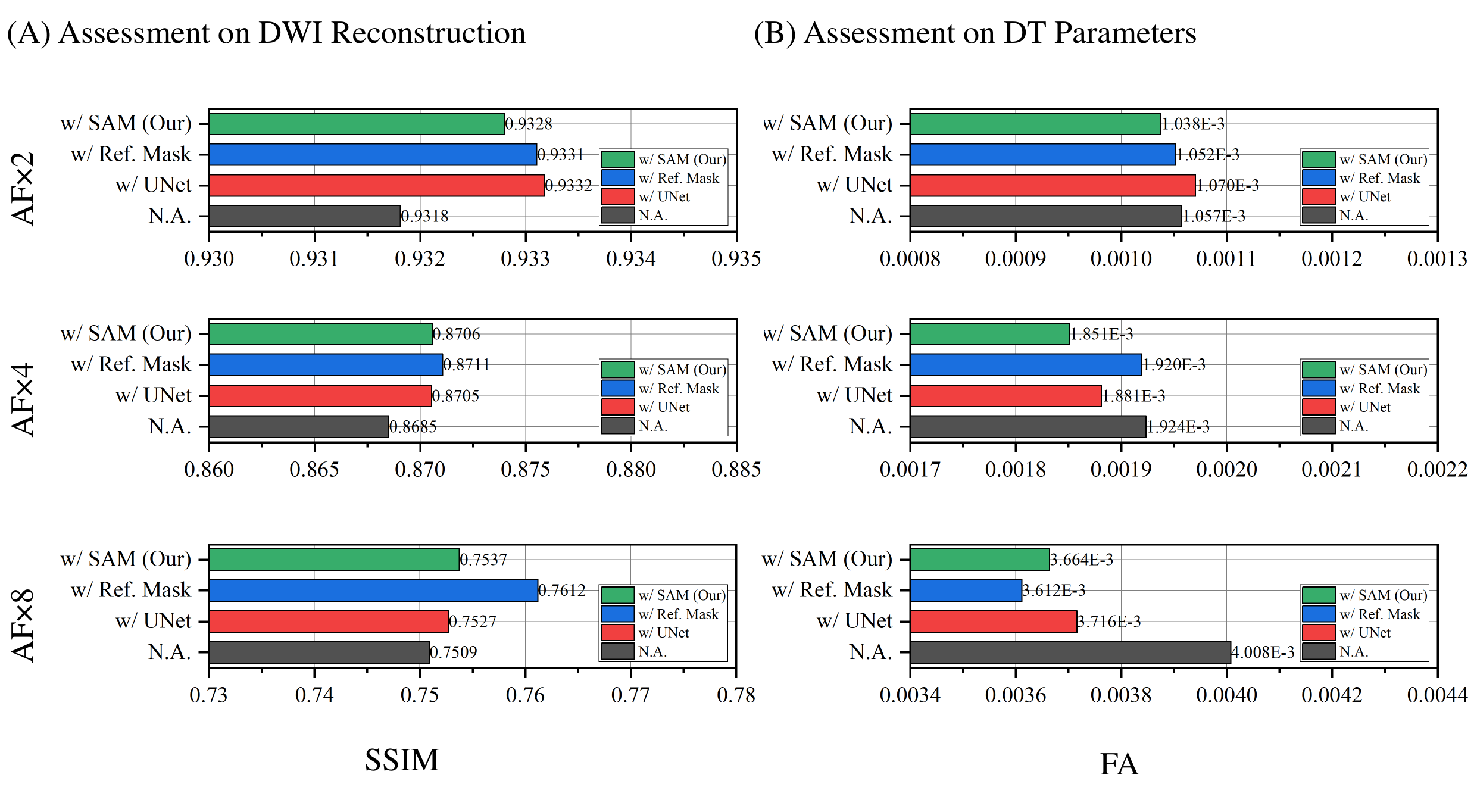}
    \caption{
    Ablation studies on the Segmentation Model using SSIM and MAE of global FA.
    w/ SAM: Our proposed RSFR; 
    w/ Ref. Mask: Replace the Segmentation Model with the manual annotation;
    w/ UNet: A pre-trained UNet for myocardium segmentation;
    N.A.: Remove the Segmentation Model.
    }
    \label{fig:FIG_ABL_SEG}
\end{subfigure}
\begin{subfigure}
    \centering
    \includegraphics[width=\linewidth]{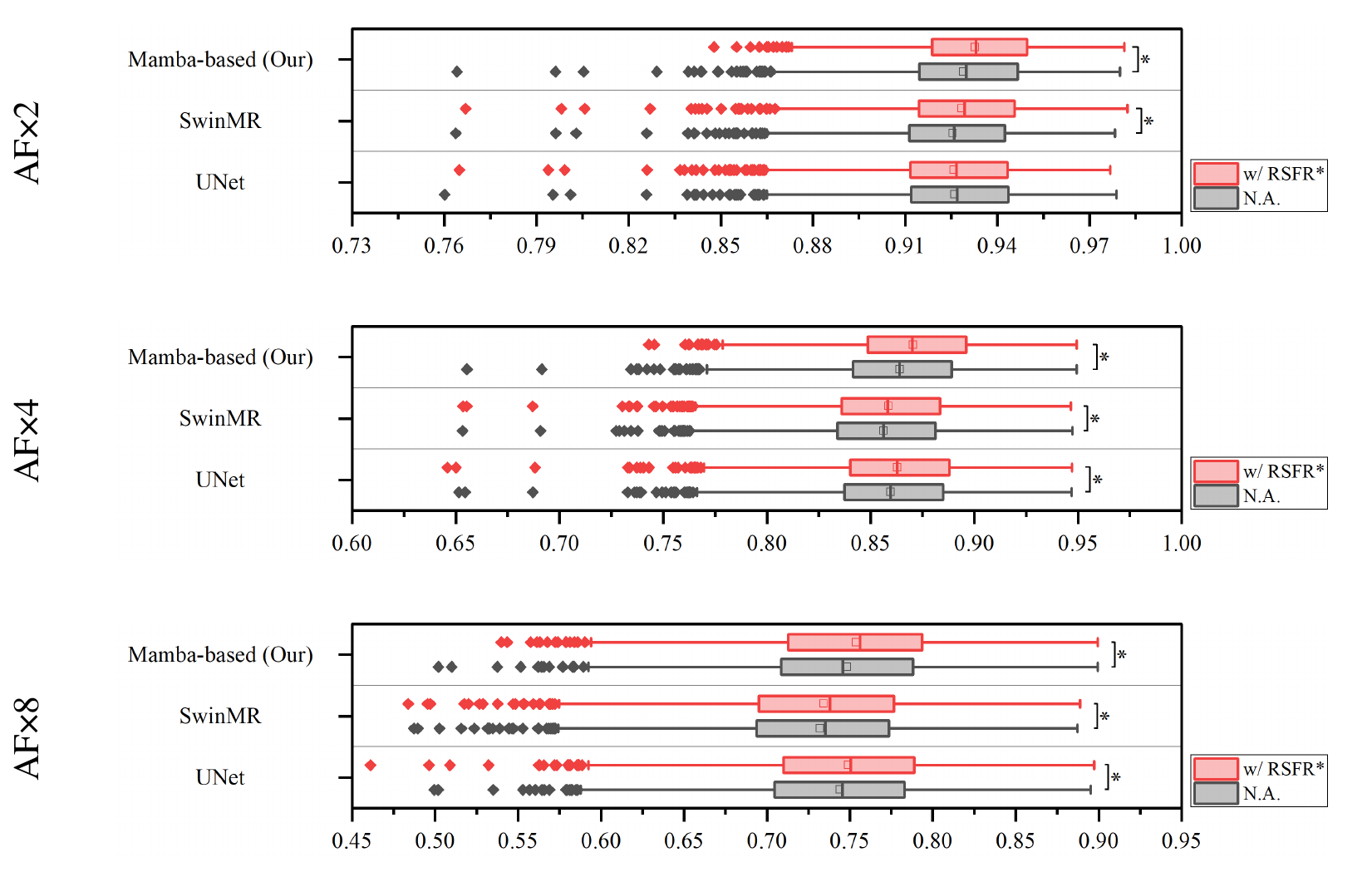}
    \caption{
    Ablation studies on reconstruction backbone model using SSIM, across Mamba-based backbone, SwinMR and UNet.
    w/ RSFR$^{*}$: Using the proposed RSFR framework; 
    N.A.: Using only the backbone;
    Our proposed RSFR Mamba-based + w/ RSFR$^{*}$.
    }
    \label{fig:FIG_ABL_RECON}
\end{subfigure}
\end{figure}

As shown in Fig.~\ref{fig:FIG_ABL_RECON}, the Mamba-based backbone with RSFR achieves the highest SSIM across all AFs, demonstrating superior robustness and reconstruction quality. 
UNet with RSFR follows closely, but shows increased variability, particularly at higher AFs. 
SwinMR with RSFR performs well at AF$\times 2$ but declines notably at AF$\times 4$ and $\times 8$, indicating sensitivity to severe undersampling. Without RSFR, all backbones show relatively suboptimal performance, with Mamba-based backbone consistently outperforming SwinMR and UNet. These results confirm that Mamba-based backbone with RSFR framework provides the best balance of robustness and quality, especially under challenging undersampling conditions.

\section{Discussion}
\label{sec:discussion}

% Reclaim the Contribution
In this work, we have proposed a novel cardiac DWI reconstruction framework, RSFR, for cDTI.
The RSFR framework was primarily motivated by the critical limitations of current cDTI reconstruction methods, particularly in addressing challenges associated with low SNR and the severe artefacts observed in undersampled cardiac DWIs. Experimental findings have validated the effectiveness of our proposed solutions ({S1}-{S3}) in addressing all challenges ({Q1}-{Q3}) outlined in Fig.~\ref{fig:FIG_INTRO}.

% Comparison Study
% - Quantitave Table (DWI)
Comparison experiments were conducted against baseline and SOTA methods, including D5C5~\cite{Schlemper2017DCCNN}, DAGAN~\cite{Yang2018DAGAN}, STGAN~\cite{Huang2022STGAN}, UNet~\cite{Ronneberger2015UNet}, SwinMR~\cite{Huang2022SwinMR}, and MambaMIR~\cite{Huang2024Enhancing_MambaMIR}.
In terms of metrics focusing more on reconstruction fidelity (PSNR and SSIM), the proposed RSFR consistently outperformed all other methods across all undersampling rates (AF$\times 2$, AF$\times 4$, AF$\times 8$), as presented in Table~\ref{tab:TAB_COMP}. 
Notably, at the extremely high undersampling rate of AF$\times 8$, where reconstructing high-fidelity images from sparse \textit{k}-space data is particularly challenging, RSFR achieved the highest SSIM (0.754) and PSNR (24.25), demonstrating its superior stability and powerful reconstruction capability compared to the other methods.
In terms of perceptual quality metrics (LPIPS), generative model-based methods (DAGAN, STGAN) tended to produce results with better perceptual fidelity, consistent with observations in previous MRI reconstruction studies. 

% - Visualised Samples (DWI)
The reconstructed DWIs in Fig.~\ref{fig:FIG_RECON_VIS} qualitatively compare the performance of RSFR with competing methods.
% - - AF2
At AF$\times 2$, most models, including UNet and SwinMR, produced visually acceptable reconstructions but exhibited subtle noise and artefacts in the myocardial region. RSFR, in contrast, delivered reconstructions with minimal artefacts and well-preserved myocardial boundaries.
% - - AF4
At AF$\times 4$, the differences became more pronounced. Models like D5C5 and DAGAN struggled to reconstruct myocardial structures, while UNet and SwinMR exhibited blurring and loss of fine details. RSFR maintained sharper myocardium reconstructions (green arrow, Fig.~\ref{fig:FIG_RECON_VIS} row 4), effectively mitigating aliasing artefacts and background noise.
% - - AF8
At the most challenging AF$\times 8$, the advantages of RSFR were even more evident. 
Competing methods suffered severe degradation, with DAGAN failing to maintain myocardial structure due to substantial noise, D5C5 introducing significant artefacts, and STGAN demonstrating poor model convergence. UNet struggled with structural preservation and high residual errors within the myocardial ROI, while SwinMR performed slightly better but showed blurring and limited detail retention. MambaMIR preserved structure better than other competing methods, but introduced artefacts near myocardial boundaries and higher residual errors.
In contrast, RSFR excelled at AF$\times 8$, producing sharp myocardial walls (red arrow, Fig.~\ref{fig:FIG_RECON_VIS} row 6) with minimal artefacts, the lowest residual errors, and effective noise suppression.

% - Quantitative Boxplot (DT)

In terms of DT parameters quality, RSFR consistently achieved the lowest MAE, highlighting its superior quantitative accuracy and structural detail preservation.
Notably, the smaller interquartile ranges of RSFR reflected its high consistency and robustness, particularly for FA, where it outperformed all methods by a clear margin. While MambaMIR showed competitive performance, it exhibits slightly higher variability. SwinMR and UNet struggled with higher MAE values, especially at high AFs, underscoring RSFR's superior robustness and accuracy.

% - Visualised DT Maps FA MD (DT)
Fig.~\ref{fig:FIG_DT_VIS} presents the FA and MD maps, along with their corresponding error maps, at AF$\times 4$. 
RSFR achieved the lowest MAE values, demonstrating its superior capability in preserving both quantitative accuracy and structural fidelity. The FA and MD maps generated by RSFR closely aligned with the reference, exhibiting minimal artefacts and noise. 
MambaMIR achieved competitive MAE values, with its maps reasonably matching the reference; however, minor artefacts in the myocardial region resulted in slightly elevated error levels. 
UNet produced higher MAE values, with its FA and MD maps showing noticeable deviations from the reference. The corresponding error maps revealed greater residual errors concentrated in the myocardial ROI.
SwinMR performed the worst among the methods, with the highest MAE values, pronounced artefacts and distortions in the maps.

% - Visualised DT Maps HA (DT)
% basic
Fig.~\ref{fig:FIG_DT_VIS_MI_ANALYSIS} illustrates the HA maps and HA line profiles with linear regression for an MI case, comparing the proposed RSFR method against UNet, SwinMR, and MambaMIR at AF$\times 4$ and $\times 8$.
% reason
In healthy myocardium, the HA typically exhibits a gradual, linear orientation change of myocardial fibres, transitioning from left-handed near the endocardium to right-handed near the epicardium. In MI cases, however, infarcted regions often show a significant reduction in right-handed myocytes, disrupting this normal transmural progression. This pathological remodelling is reflected in the abnormal distributions observed in HA maps~\cite{Sharrack2022Relationship}.
% results
The results demonstrate the superior ability of RSFR to maintain both structural and quantitative fidelity, even in pathologically altered myocardium.
% AF4
At AF$\times 4$, all methods recovered the general HA distribution to varying extents, with RSFR showing the best alignment to the reference. The HA LP revealed that RSFR maintains a closer fit to the reference, achieving high $R^{2}$ values and low RMSE, thereby accurately capturing the transmural gradient of HA.
% AF8
At AF$\times 8$, the differences between methods became stark. RSFR remained robust, closely replicating the reference HA map and line profile, including the disrupted HA progression in the MI region. Its $R^{2}$ and RMSE values remained near the reference, underscoring its ability to preserve both structural and quantitative fidelity under severe undersampling. In contrast, MambaMIR and UNet failed to accurately reconstruct the pathological region, with parts of the MI area either disappearing or being overly smoothed. SwinMR exhibited disordered HA maps and profiles, with disruptions extending beyond the MI region, leading to significant structural distortion and poor quantitative alignment with the reference, as evidenced by its lower global $R^{2}$ values and higher RMSE.
% Transition
The transition from AF$\times 4$ to AF$\times 8$ highlighted the superior consistency of RSFR in HA reconstruction within the MI region, establishing it as the most reliable choice for clinical applications, particularly where competing methods fail under higher AFs.

% Relationship between DWI recon & DT
The analysis revealed a strong correlation between DWI reconstruction fidelity and DT quality, as evidenced by metrics such as PSNR and SSIM that consistently aligned with lower global MAE values, indicating that improving DWI reconstruction fidelity enhances DT accuracy. 
In contrast, LPIPS, a perceptual metric that assesses visual similarity, showed limited relevance to DT parameter accuracy, underscoring the trade-off between perceptual similarity and pixel-wise fidelity~\cite{Blau2018Perception}.
Generative model-based methods, such as DAGAN and STGAN, often prioritise perceptual quality by minimising latent space distances~\cite{Anwar2020Deep}. However, these approaches may introduce visually appealing but clinically irrelevant ``hallucinations''\cite{Bhadra2021Hallucinations}, which, while improving perceptual scores, compromise the pixel-wise fidelity crucial for quantitative cardiac DTI\cite{Huang2024Deep}. 
Given the sensitivity of cardiac DTI to minor intensity discrepancies, prioritising pixel-wise fidelity over perceptual quality is vital to ensure accurate and reliable DT reconstructions.

% Ablation Studies
In this study, ablation experiments have underscored the critical contributions of each component within the RSFR framework, focusing specifically on segmentation models and reconstruction backbones.
In addition, the modular design of RSFR ensures flexibility and adaptability, allowing seamless integration of alternative segmentation models or reconstruction backbones. This future-proof framework can readily incorporate advancements in deep learning and MRI reconstruction, enhancing its potential for clinical translation and addressing emerging imaging challenges.

% Limitations and Future Work
Despite RSFR's SOTA performance in DWI reconstruction, there are areas for further enhancement and expansion.
% - Model Efficiency
First, the efficiency of RSFR can be improved by reducing the computational overhead of its three-subnetwork framework. This could be achieved through a lightweight reconstruction backbone or weight sharing between the two reconstruction backbones, maintaining performance while lowering complexity.
% - Extension to Other Anatomical Structures
Second, this study focused on cardiac DWI, a relatively homogeneous structure. 
Extending the proposed RSFR to more complex anatomical regions, such as the brain or abdomen, offers significant potential, as the semantic-aware approach could effectively address higher structural complexity.

\section{Conclusion}
\label{sec:conclusion}

This study presented RSFR (Reconstruction, Segmentation, Fusion \& Refinement), a novel framework for cardiac DWI reconstruction addressing challenges such as low SNR, aliasing artefacts, and quantitative fidelity. 
Leveraging a coarse-to-fine strategy, zero-shot semantic priors via SAM, and integration of semantic features, RSFR has achieved SOTA reconstruction quality and DT parameter accuracy, even under extreme undersampling conditions.
Comprehensive experiments and ablation studies have validated its effectiveness, highlighting its ability to preserve structural integrity and quantitative fidelity. With a modular and adaptable design, RSFR has represented a significant advancement in cDTI reconstruction, positioning RSFR as a significant advancement in cDTI reconstruction for potential clinical translation.

% \section*{Acknowledgments}
% This study was supported in part by the ERC IMI (101005122), the H2020 (952172), the MRC (MC/PC/21013), the Royal Society (IEC/NSFC/211235), the NVIDIA Academic Hardware Grant Program, the SABER project supported by Boehringer Ingelheim Ltd, NIHR Imperial Biomedical Research Centre (RDA01), Wellcome Leap Dynamic Resilience, UKRI guarantee funding for Horizon Europe MSCA Postdoctoral Fellowships (EP/Z002206/1), the British Heart Foundation under Grant RG/F/23/110115 and FS/19/22/34334, the Chan Zuckerberg Initiative DAF which is an advised fund of the Silicon Valley Community Foundation under Grant 2024-337787, the Engineering and Physical Sciences Research Council under Grant EP/X014010/1 and the UKRI Future Leaders Fellowship (MR/V023799/1).

%%Harvard
\bibliographystyle{model2-names.bst}\biboptions{authoryear}
\bibliography{refs}

% \section*{Supplementary Material}

\end{document}